\input ./mtexsis.tex
\texsis
\paper\tenpoint
\superrefsfalse 
\hfuzz 2.cm
\font\sstenz=cmss10 scaled \magstep0


\def\frac#1#2{{#1\over#2}}
\def\As{A\!\!\!\slash\,}

\def\ps{p\!\!\!\slash\,}


\def\scp#1#2{\langle{#1}\vert{#2}\rangle}
\def\ah{{\hat{a}}}

\def\bra#1{\langle{#1}\vert}
\def\ket#1{\vert{#1}\rangle}

\def\ad{{\hat{a}}^\dagger}




\def\n{\hfill\break}
\def\frac#1#2{{#1\over#2}}
\def\square{\hbox{{$\sqcup$}\llap{$\sqcap$}}}%
%

\def\part{\partial}



\referencelist
\reference{[Garrod]} 
G.~Garrod, Phys. Rev. {\bf 167}, 1143 (1968)
\endreference

\reference{[HG]}
H.~Goldstein, {\it Classical Mechanics} 
(Addison Wesley, Reading Mass., 1959) p. 207
\endreference

\reference{[LLQED]}
V.~B.~Berestetskii, E.~M.~Lifshitz, and L.~P.~Pitaevskii, {\it Quantum Electrodynamics, Second Edition} 
(Pergamon Press, New York, 1982)
\endreference

\reference{[SandK]}
G.~Stephenson and C.~W.~Kilmister, {\it Special Relativity for Physicists} 
(Longmans, Green and Co., London, 1958) p. 65
\endreference

\reference{[TGCC]}
T.~Garavaglia, in
\booktitle{Proceedings of the 1991 Symposium on the 
Superconducting Super Collider, Corpus Christi, SSCL-SR-1213},
edited by V.~Kelly and G.~P.~Yost
(Superconducting Super Collider Laboratory, Dallas Texas, 1991) p. 669
\endreference

\reference{[TGSF]}
T.~Garavaglia, in
 {\it Conference Record of the 1991 IEEE Particle Accelerator Conference, 
San Francisco}, edited by L.~Lizama and J.~Chew (IEEE, 1991) 
Vol. I, p. 231,\n 
(See also http://WWW.hep/ftp/ssc/new/COLLIDER/colliderpapers, SSCL-416 Rev)
\endreference

\reference{[Fock]} 
V.~Fock, Physik Z. Sowjetunion {\bf 12}, 404 (1937) 
\endreference

\reference{[Schwinger]}
J.~Schwinger, Phys. Rev. {\bf 82}, 664 (1951)
\endreference

\reference{[WIGTR]} 
E.~P.~Wigner, Nachrichten d. Gesell. d. Wissenshaften z. 
G\"ottingen Math.-Phys. Klass, 546 (1932)
\endreference

\reference{[WIGAU]} 
E.~P.~Wigner, J. Math. Phys. {\bf 1}, 409 (1960); 
J. Math. Phys. {\bf 1}, 414 (1960)
\endreference

\reference{[TG-SK]} 
T.~Garavaglia and S.~K.~Kauffmann, {\it The principle of symmetric bracket
invariance as the origin of first and second quantization}, 
(http://xxx.lanl.gov, hep-th/9907059), 
Contributed paper to XIX International Symposium on Lepton
and Photon Interactions at High Energies, 
Stanford University, August 9-14, 1999.(http://lp99.slac.stanford.edu)
\endreference

\reference{[WP]}
W.~Pauli, {\it Prinzipien der Quantentheorie 1, Handbuch der Physik, Band V, Teil 1.} 
(Springer-Verlag, Berlin, 1958) p. 60
\endreference

\reference{[PandW]} 
W.~Pauli, and V.~Weisskopf, Helv. Phys. Acta {\bf 7}, 709(1934)
\endreference

\reference{[Dirac33]} 
P.~A.~M.~Dirac, Physikalische Zeitschrift der Sowjetunion, Band {\bf 3},
Heft {\bf 1}, 64 (1933)
\endreference

\reference{[GandR]}
I.~S.~Gradshteyn and I.~M.~Ryzhik, {\it  Tables of Integrals, Series and
Products, Fifth Edition} 
(Academic Press, New York, 1996) 
\endreference

\reference{[A-WIG]} 
S.~M.~W.~ Ahmad and E.~P.~Wigner, Il Nuovo Cimento {\bf 28A}, 1 (1975)
\endreference

\reference{[OC-WIG]} 
R.~F.~O'Connell and E.~P.~Wigner, Phys. Lett. {\bf 61A}, 353 (1977); 
Phys. Lett. {\bf 67A}, 319 (1978)
\endreference

\endreferencelist

\line{{covqmp.tex\hfill dias-stp-00-22}}
\n\n
\centerline{\bf Covariant relativistic quantum theory}
\n
\centerline{T. Garavaglia$^{*}$}
\n
\centerline{\it Institi\'uid \'Ard-l\'einn Bhaile \'Atha Cliath, Baile
\'Atha Cliath  4, \'Eire$^{\dagger}$}
\vskip 24pt
\par
\vbox{\sstenz 
\par
Covariant classical particle dynamics is described, and the associated
covariant relativistic particle quantum mechanics is derived. 
The invariant symmetric bracket is defined on the space of
quantum amplitudes, and its relation to a generalized
Hamiltonian dynamics and to a covariant Schr\"odinger type
equation is shown. 
Examples for
relativistic potential problems are solved.  Mathematically and
physically acceptable probability densities for the Klein-Gordon equation
and for the Dirac equation are derived, and a new continuity equation for
each case is given. 
The quantum distribution for mass is discussed, and  
unambiguous representations of four-velocity and four-acceleration
operators are given.} 
 
\vskip 20pt 
\vskip 24pt
\hrule \n
\item{$^*$}{E-mail: bronco@stp.dias.ie }  
\item{$^\dagger$}{Also {\it Institi\'uid Teicneola\'iochta Bhaile 
\'Atha Cliath.}}
\vfill
\supereject
\section{INTRODUCTION}
\par
The covariant form of classical dynamics \cite{[Garrod]} and \cite{[HG]}
provides an elegant method for analyzing problems in relativistic particle
dynamics. This is briefly described, and the corresponding
covariant quantum formulation is presented.
The method of analyzing potential problems in both the classical and the
quantum case is given. Among the new results that follow from this analysis
are the resolution of the ambiguity associated with the interpretation of
probability density for the Klein-Gordon equation, a new equation of
continuity for this case, and a  similar equation for the Dirac equation.
Also a representation of the covariant four-velocity operator is given.
Since the covariant Hamiltonian operator is quadratic in four-momentum in 
the scalar case and
linear in the spinor case, this four-velocity operator is unambiguously
defined, and it is related to the corresponding classical quantities through the 
expectation values of the quantum operator.
In addition, an exact quantum solution is found  for the
relativistic motion of a spin zero particle in a uniform field. The method
can be easily extended to more complicated examples.
\par
The Lorenz invariant parameter that is relevant for dynamical development
in classical covariant mechanics is proper distance $s$, proper time times the
speed of light, and the invariant parameter that characterizes the rest
energy of a particle is its mass. 
These quantities have a fundamental quantum nature, and quantum distributions 
are associated
with them. The relation of the c-number values of these quantities, which
appear in the quantum and classical equations, to associated 
operators is described. 
The evolution of
quantum states with respect to the  parameters $s$ is investigated. 
\par 
The covariant free particle Green's function is derived, and it is shown
how it provides a description of the evolution of initial quantum states. It
is seen how an initial Gaussian state, in which the space-time coordinates
are treated on an equal footing, spreads with respect to the proper time 
evolution parameter about the classical covariant path. 
Furthermore, the evolution of
completely localized and completely delocalized initial states is discussed.
It is interesting to observe how closely the covariant relativistic quantum
results parallel the corresponding  non-relativistic quantum results.
\par 
The covariant Schr\"odinger type equation describes the evolution of the
quantum state for a massive scalar particle or spinor particle. The
probability densities for observables are found from the quantum amplitudes
associated with these states. The notion of the symmetric bracket defined on
the space of quantum amplitudes is introduced, and it is shown how it is
related to the covariant Schr\"odinger type equation. This results in the
invariance of this bracket with respect to an one parameter transformation
induced by a generator, which is bilinear in the quantum amplitudes. The
coefficients of these amplitudes are the matrix elements of the covariant
Hamiltonian that appears in either the scalar or the spinor covariant
Schr\"odinger type
equation. It is shown how this generator plays the role of an Hamiltonian
on a generalized phase space formed from the real and imaginary parts of
these amplitudes. The Hamiltonian flow generated on this space corresponds
to the quantum state evolution provided by the covariant Schr\"odinger type
equation.
\par
The conventions used for four-vector, spinors,  and Dirac gamma matrices are those of
\cite{[LLQED]} with $\hbar=c=1$.
The space-time contravariant coordinate four-vector is defined as
$q^\mu=(q^0,q^1,q^2,q^3)=(ct,x,y,z),$
and the dimensionless form is found by dividing the components with the 
fundamental units of length $q_s$. The Einstein summation convention is used
throughout, and the covariant components are found from
$q_\mu=g_{\mu\nu}q^\nu$,
with the non-zero components of the metric tensor given by
$(g_{00},g_{11},g_{22},g_{33})=(1,-1,-1,-1).$
The scalar product of two four-vectors is $q\cdot p=q_\mu p^\mu$, and
the space-time measure is
$ds=\sqrt{dq\cdot dq}$.
The four-velocity is defined as
$u^\mu={{dq^\mu}/{ds}}$, 
and $u_\mu u^\mu=1$.
The four-momentum for a free particle is  
$p^\mu=mu^\mu=(p^0,\vec p)=(\gamma m,\gamma{\vec \beta}m)$,
with ${\vec \beta}={{d{\vec q}}/{dq^0}}$,  
$\gamma=1/\sqrt{1-{\vec \beta}\cdot{\vec \beta}}$, 
and $p\cdot p=m^2$.
\par
The square root of the space-time measure can be found in another way, which
leads to the linear representation of the relativistic wave equation
found by Dirac. This method is based on the introduction of the four
matrices $\gamma^\mu$,\,($\mu=0,1,2,3$), which satisfy the relation 
$$
\gamma^\mu\gamma^\nu+\gamma^\nu\gamma^\mu=2g^{\mu\nu}I.
\EQN{qms.6}$$
In this way the space-time measure has the representation
$d\slashchar{s}=\gamma\cdot dq$, and  $ds^2=d\slashchar{s}\cdot
d\slashchar{s}={dq\cdot dq}$.

\section{COVARIANT ACTION AND COVARIANT CLASSICAL DYNAMICS}

\par
The covariant classical methods, which are necessary for the development of
covariant quantum theory, are described here.
The covariant action is defined as
$$
{\cal S}=\int {\cal L}(q,u)ds,
\EQN{ac.1}$$
where ${\cal L}(q,u)$ is a Lorentz invariant.
The requirement that the action be an extremum under a variation leads to the 
covariant Euler-Lagrange equation
$$
\frac{d \frac{\partial {\cal L}}{\partial u^\mu}}{ds}
-\frac{\partial {\cal L}}{\partial q^\mu}=0.
\EQN{ac.2}$$
\par
The covariant Lagrangian for a particle of mass $m$ and charge $e$ interacting 
with a four-vector field $A^\mu(q)$ is 
$$
{\cal L}=\frac{m}{2}u^2-eA\cdot u,
\EQN{ac.3}$$
and the generalized four-momentum is
$$
p_\mu=\frac{\partial {\cal L}}{\partial u^\mu}=mu_\mu-eA_\mu.
\EQN{ac.4}$$
The equation of motion is found from \Eq{ac.2} to be
$$
\frac{dp_\mu}{ds}+\partial_\mu{eA\cdot u}=0,
\EQN{ac.5}$$
where $\partial_\nu=\partial/\partial q^\nu$.
Since
$$
\frac{d A_\mu(q)}{ds}=u^\nu\partial_\nu{A_\mu(q)},
\EQN{ac.6}$$
the equation of motion becomes
$$
ma_\mu=eu^\nu F_{\nu\mu},
\EQN{ac.7}$$
with
$$
a_\mu=\frac{d^2 q_\mu}{ds^2},\quad 
F_{\nu\mu}=\partial_\nu A_\mu-\partial_\mu A_\nu.
\EQN{ac.8}$$


\par
From the covariant Lagrangian \Eq{ac.3}, the covariant Hamiltonian is defined 
as 
$$
{\cal H}=p\cdot u-{\cal L}.
\EQN{chd.1}$$
Since $u_\mu=(p_\mu+eA_\mu)/m$, a simple calculation gives
$$
{\cal H}=\frac{(p+eA)^2}{2m}=\frac{(mu)^2}{2m}=\frac{m}{2}.
\EQN{chd.2}$$
From \Eq{chd.1} and
$$
\frac{d{\cal L}}{ds}=\frac{\partial{\cal L}}{\partial q}\cdot u
+\frac{\partial{\cal L}}{\partial u}\cdot a,
\EQN{chd.3}$$
it follows that
$$
\frac{d{\cal H}}{ds}=(\frac{d p}{ds}-\frac{\partial{\cal L}}{\partial q})\cdot u
-(p-\frac{\partial{\cal L}}{\partial u})\cdot a=0,
\EQN{chd.4}$$
which is a consequence of \Eq{ac.2}.
In addition
$$
\frac{d{\cal H}}{ds}=\frac{\partial{\cal H}}{\partial q}\cdot u
+\frac{\partial{\cal H}}{\partial p}\cdot f=0,
\EQN{chd.5}$$
with $f^\nu=d p^\nu/ds$, and this implies the Hamilton equations
$$
u_\mu=\frac{\partial{\cal H}}{\partial p^\mu}
\qquad f_\mu=-\frac{\partial{\cal H}}{\partial q^\mu}.
\EQN{chd.6}$$
These equations and the covariant Hamiltonian \Eq{chd.2} give
$$
u_\mu=\frac{\partial{\cal H}}{\partial p^\mu}=(p+eA)_\mu/m,
\EQN{chd.7}$$
and
$$
f_\mu=-\frac{\partial{\cal H}}{\partial q^\mu}
=-u^\nu\partial_\mu eA_\nu.
\EQN{chd.8}$$
From \Eq{chd.7} it follows that
$$
f_\mu=ma_\mu-e\frac{d A_\mu}{ds},
\EQN{chd.9}$$
which with \Eq{ac.6} and \Eq{chd.8} results in the equation of motion \Eq{ac.7}.


\par
For real functions $F(q,p,s)$ and $G(q,p,s)$ of $q_\nu$, $p_\nu$, and $s$, 
the Poisson bracket is defined as
$$
\{F,G\}=\partial F\cdot {\bar \partial}G-\partial G\cdot {\bar \partial}F,
\EQN{pb.1}$$
with $\partial_\mu=\partial/\partial q^\mu$ and 
${\bar \partial}_\mu=\partial/\partial p^\mu$.
Since 
$$
\partial_\mu q^\nu= {\bar\partial}_\mu p^\nu=\delta_\mu^\nu,
\EQN{pb.2}$$
one finds
$$
\{q^\mu,p^\nu\}=g^{\mu\nu}.
\EQN{pb.3}$$
From
$$
\frac{dF}{ds}
=\frac{\partial F}{\partial q}\cdot u
+\frac{\partial F}{\partial p}\cdot f+\frac{\partial F}{\partial s},
\EQN{pb.4}$$
and \Eq{chd.6}, it follows that
$$
\frac{d F}{ds}=\{F,{\cal H}\}+\frac{\partial F}{\partial s}.
\EQN{pb.5}$$
The dynamical equations that are equivalent to Hamilton's equations 
\Eq{chd.6} are
$$
u_\mu=\{q_\mu,{\cal H} \}\qquad f_\mu=\{p_\mu,{\cal H} \}
\EQN{pb.6}$$
\par
The Hamiltonian is the generator for a one parameter transformation that
leaves the Poisson bracket invariant. This is seen from
$$
\frac{d}{ds}\{q(s)^\mu,p(s)^\nu\}=\{\frac{dq(s)^\mu}{ds},p(s)^\nu\}+
\{q(s)^\mu,\frac{dp(s)^\nu}{ds}\}=0,
\EQN{pb.7}$$
which follows from the dynamical equations \Eq{pb.6} and the Jacobi identity
for the Poisson bracket. This leads to the conclusion
that 
$$
\{q(s)^\mu,p(s)^\nu\}=\{q(0)^\mu,p(0)^\nu\}=g^{\mu\nu}.
\EQN{pb.8}$$


\par
To illustrate the use of the covariant classical dynamical equations,
examples are given that are to be compared with the
corresponding covariant quantum solutions. These are associated with
a vector potential of the form $A^\mu=(A^0,\vec A)=(-V(\vec q),0)$.
From the covariant Hamiltonian \Eq{chd.2} and the dynamical equations
\Eq{chd.6} or \Eq{pb.6}, it follows that
$$
u^0=(p^0+A^0)/m\qquad f^0=\frac{(p+A)^0}{m}\partial_0 A_0=0;
\EQN{pp.1}$$
hence, $p^0={E}$, a constant, and
$$
{E}=m\gamma+V(\vec q),
\EQN{pp.2}$$
since $u^0=dq^0/ds=\gamma$. Also
$$
f^i=\frac{(p+A)^0}{m}\partial_i A_0=-u^0\partial_i V(\vec q)
\EQN{pp.3}$$
and
$$
\frac{d(m\gamma\vec \beta)}{dq^0}=-\grad V(\vec q). 
\EQN{pp.4}$$
For a free particle with $V(\vec q)=0$, one finds $E=\gamma m$, 
and $\vec p=\gamma m\vec \beta$. 
\par
For a particle in a uniform field with potential
$V(z)=-Kz,$
the dynamical equations give
$$
\frac{dq^0}{ds}=(p^0-V(z))/m, \quad \frac{dp^0}{ds}=0,\quad p^0={E},
\EQN{pp.6}$$
and
$$
\frac{dq^3}{ds}=\frac{p^3}{m},\quad
\frac{dp^3}{ds}=({E}-V(z))\frac{\partial V(z)}{\partial z}.
\EQN{pp.7}$$
Since $p^3=m\dot z(dq^0/ds)=m\dot z({E}-V(z))/m$, the invariant Hamiltonian
becomes
$$
({E}-V(z))^2(1-{\dot z}^2)=m^2.
\EQN{pp.8}$$
For $z(0)= \dot z(0)=0$, the solution is
$$
z(t)=\frac{m}{K}(\sqrt{1+(\frac{Kt}{m})^2} -1).
\EQN{pp.9}$$
Additional related
relativistic particle dynamics is found in \cite{[SandK]}, and applications of 
covariant dynamics in curvilinear coordinates to betatron
physics may be found in \cite{[TGCC]} and \cite{[TGSF]}. Related concepts of
covariant dynamics have been used also in the development of relativistic 
wave equations \cite{[Fock]} and covariant
quantum field theory \cite{[Schwinger]}.
\par

\section{COVARIANT QUANTUM MECHANICS AND SYMMETRIC BRACKET INVARIANCE}

\par It is the objective of this paper to establish a quantum theory that
corresponds to the covariant classical theory described above, where the 
dynamical parameter is proper time $s$. As a
starting point, the scalar case is considered, and the extension to the
spinor case is given later. This state is normalized to unity, a Lorentz invariant, and for the probability interpretation of
quantum theory, it is to remain normalized to this value for changes in the
value of the parameter $s$. This is represented as
$$
\scp{\Psi(s)}{\Psi(s)}=\scp{\Psi(0)}{\Psi(0)}=1,
\EQN cqmsb.1$$
and this requires the state $\Psi(s)$ to be related to the original state
$\Psi(0)$ by either a unitary or antiunitary transformation \cite{[WIGTR]}
and \cite{[WIGAU]}.
It is well known that the antiunitary transformation is associated with the
$TCP$ theorem \cite{[LLQED]}, and the norm preserving transformation
associated with conservation of probability is unitary. 
The proper time developed state is 
induced by a one parameter unitary transformation ${\hat U}(s)$ such
that
$$
\ket{\Psi(s)}=\hat U(s)\ket{\Psi(0)}.
\EQN cqmsb.2$$
The product of two such transformations is of the form
$$
{\hat U}(s_1){\hat U}(s_2)={\hat U}(s_1+s_2),
\EQN cqmsb.3$$
with
$$
{\hat U}(s)=e^{-is{\hat {\cal H}}},
\EQN cqmsb.4$$
where ${\hat {\cal H}}$ is an Hermitian operator. The differential
representation of \Eq{cqmsb.1} is the covariant Schr\"odinger type 
equation
$$
i{\partial \ket{\Psi(s)}\over\partial s}=\hat{\cal H}\ket{\Psi(s)}.
\EQN cqmsb.5$$
Under a Lorentz transformation, the state $\ket{\Psi(s)}$  changes
according to $\ket{\Psi'(s)}=\Lambda(\vec\beta)\ket{\Psi(s)}$, where 
$\Lambda(\vec\beta)$ is a scalar
representation of the Lorentz group, and $\vec\beta$ is a dimensionless 
velocity vector. 
Under this transformation the
norm is invariant and so is the form of \Eq{cqmsb.5}.
\par The spinor case requires some modifications of the above discussion.
A general spinor state $\ket{\Psi(0)}$ has associated with it a conjugate
state found from the adjoint state and represented as
$$
\bra{\bar\Psi(0)}=\bra{\Psi(0)}\gamma^0,
\EQN cqmsb.6$$
where $\gamma^0$ is the time component Dirac matrix.
In this case the scalar product is normalized to unity according to
$$
\scp{\bar\Psi(s)}{\Psi(s)}
=\scp{\bar\Psi(0)}{\Psi(0)}=1.
\EQN cqmsb.7$$
This scalar product is a Lorentz invariant when the states transform
according to
$$
\ket{\Psi'(s)}=\Lambda(\vec\beta)\ket{\Psi(s)}\quad 
\bra{\bar\Psi'(s)}=\bra{\bar\Psi(s)}\Lambda^{-1}(\vec\beta),
\EQN cqmsb.8$$ 
where 
$\Lambda(\vec\beta)$ is a spinor representation of the Lorentz group,
and
$$
\Lambda^{-1}(\vec\beta)=\gamma^0\Lambda(\vec\beta)\gamma^0.
\EQN cqmsb.9$$
The evolution of the spinor state $\ket{\Psi(0)}$ is generated by the
transformation $\hat U(s)$, which has the inverse
$$
\hat U^{-1}(s)=\gamma^0\hat U^\dagger(s)\gamma^0,
\EQN cqmsb.10$$
which preserves the scalar product \Eq{cqmsb.7}.
\par

\par For the scalar state case, an eigenstate in the rest system of a particle of mass $m$
is represented by $\ket{\lambda_i}$, where $\lambda$ represents
a set of observables associated with measurements of the state.
The general state of the particle is expanded at $s=0$ in terms of 
these states, and the evolution of the state is generated by $\hat U(s)$.
Introducing the identity operator, it is found that
$$
\ket{\Psi(s)}=\hat I\ket{\Psi(s)}=\sum_i\ket{\lambda_i}\scp{\lambda_i}{\Psi(s)}
=\sum_i a_i(s)\ket{\lambda_i},
\EQN cqmsb.11$$
where 
$$
a_i(s)=\scp{\lambda_i}{\Psi(s)}
=\bra{\lambda_i}\hat U(s)\ket{\Psi(0)}.
\EQN cqmsb.12$$
Differentiating with respect to $s$, it is found that
$$
i\frac{da_i(s)}{ds}=\bra{\lambda_i}\hat {\cal H}\ket{\Psi(s)}
$$
$$
=\sum_j\bra{\lambda_i}\hat{\cal H}\ket{\lambda_j}\scp{\lambda_j}{\Psi(s)}
$$
$$
=\sum_j{\cal H}_{ij}a_j(s),
\EQN cqmsb.13$$
where ${\cal H}_{ij}=\bra{\lambda_i}\hat {\cal H}\ket{\lambda_j}$.
\par
It is now shown that this equation leads to invariance of 
the symmetric
bracket defined on the space of complex quantum amplitudes $a_i$. 
For real functions $U(a_i,a^*_j)$ and  $V(a_i,a^*_j)$, 
the symmetric bracket is defined as
$$
\{U,V\}_+
=\sum_i(\frac{\partial U}{\partial a_i}\frac{\partial V}{\partial a^*_i}
+\frac{\partial V}{\partial a_i}\frac{\partial U}{\partial a^*_i}).
\EQN cqmsb.14$$
The complex coefficients $a_i$ and $a^*_j$ satisfy, in the discrete case, 
the conditions
$$
\{a_i,a_{j}\}_+=\{a^*_i,a^*_{j}\}_+=0,\qquad\{a_i,a^*_{j}\}_+=\delta_{i,j},
\EQN cqmsb.15$$
and in the continuum case these become
$$
\{a_\lambda,a_{\lambda'}\}_+=\{a^*_\lambda,a^*_{\lambda'}\}_+=0,
\qquad\{a_\lambda,a^*_{\lambda'}\}_+=\delta(\lambda-\lambda'),
\EQN cqmsb.16$$
with 
$$
\frac{\partial a_\lambda}{\partial a_{\lambda'}}=\delta(\lambda-\lambda'),
\EQN cqmsb.17$$
and its complex conjugate. Introducing the real function defined from the Hermitian matrix
${\cal H}_{ij}$
$$
g(s)=\sum_ia^*_{i}(s){\cal H}_{ij}a_j(s),
\EQN cqmsb.18$$
it follows that
$$
i\frac{da_i(s)}{ds}=\{a_i(s),g(s)\}_+=\frac{\partial g(s)}{\partial a_i^*(s)}
$$
$$
-i\frac{da^*_i(s)}{ds}
=\{a^*_i(s),g(s)\}_+=\frac{\partial g(s)}{\partial a_i(s)}.
\EQN cqmsb.19$$
Defining $a_i=a_i(0)$, the symmetric bracket invariance
follows from
$$
\{a_i(ds),a_j^*(ds)\}_+=\{a_i,a_j^*\}_+ +O(ds^2)\sim \delta_{ij}.
\EQN cqmsb.20$$
Following the methods of \cite{[TG-SK]}, the iteration of \Eq{cqmsb.19} 
shows that the development 
of $a(0)$ is generated by the operator $\hat V(s)$ defined in
$$
a_i(s)=\hat V(s)a_i(0)=e^{is\delta_+}a_i(0),
$$
which can be used to show that
$$
\{a_i(s),a_j^*(s)\}_+=\{a_i,a_j^*\}_+=\delta_{ij}.
\EQN cqmsb.21$$ 
In the definition of $\hat V(s)$, the operator $\delta_+=\{g(0),\,\}_+$ has
the properties
$$
\delta_+a_i=\{g(0),a_i\}_+, \quad
\delta_+^2a_i
=\{g(0),\{g(0),a_i\}_+\}_+,\quad etc.
\EQN cqmsb.22$$
The invariance of this bracket can be viewed as a principle, which leads to 
first and second quantization. This is discussed in detail in \cite{[TG-SK]}.
\par
It is interesting to note that the equations for the development of $a_i(s)$, \Eq{cqmsb.19},  
are equivalent to generalized
Hamilton's equations on a space where the real and imaginary parts of
the amplitude $a_i(s)=\scp{\lambda_i}{\Psi(s)}$ behave like phase space
coordinates. To show this, the coordinates $Q_i(s)$ and $P_i(s)$
are defined as
$$
Q_i(s)=\frac{a_i(s)+a_i^*(s)}{\sqrt{2}},\quad
P_i(s)=\frac{a_i(s)-a_i^*(s)}{i\sqrt{2}}.
\EQN cqmsb.23$$
From the \Eq{cqmsb.19}, one can derive the Hamilton's equations for these
coordinates,
$$\frac{dQ_i(s)}{ds}=\frac{\partial g(s)}{\partial P_i(s)},\quad
\frac{dP_i(s)}{ds}=-\frac{\partial g(s)}{\partial Q_i(s)},
\EQN cqmsb.24$$
where $g(s)$ plays the role of the Hamiltonian on the space of these
coordinates.
This function takes a simple form if the matrix elements are diagonal, 
${\cal H}_{ij}=(\lambda_i)\delta_{ij}$, and it becomes
$$
g(s)=\sum_i\lambda_i |a_i(s)|^2=\sum_i\lambda_i \frac{Q_i^2+P_i^2}{2}.
\EQN cqmsb.25$$
It must be emphasized, that the Hamiltonian flow induced by this
generator is not associated with the dynamics of ordinary phase space
characterized by observable position coordinates $q_i$ and observable conjugate momentum 
coordinates $p_i$, but it is the dynamical flow on the space of quantum
amplitudes, $a_i(s)=\scp{\lambda_i}{\Psi(s)}$, which are not observables.
The Hamiltonian flow generated by
\Eq{cqmsb.18}
is equivalent to the covariant Schr\"odinger type equation \Eq{cqmsb.5}, 
and it is the form of \Eq{cqmsb.18} and \Eq{cqmsb.19} that accounts for 
the linearity of this equation. Furthermore, the form of the generator 
in \Eq{cqmsb.25} that leaves the symmetric bracket invariant implies the 
probability interpretation of quantum theory, where $|a_i(s)|^2$ is the
probability to observe the eigenvalue $\lambda_i$. The modification of the 
above discussion to include the spinor case is straightforward, and it follows
from the properties associated with the conjugate state \Eq{cqmsb.6} and 
the scalar product \Eq{cqmsb.7}.  

\section{COVARIANT SCHR\"ODINGER TYPE WAVE EQUATION}

\par 
The Schr\"odinger type equation in covariant form for the scalar case 
with wave function 
$$
\Psi(s,q)=\scp{q}{\Psi(s)},
\EQN cv.1$$
is
$$
i{\partial \Psi(s,q)\over\partial s}=\hat{\cal H}\Psi(s,q).
\EQN cv.2$$
Using a separable solution $\Psi(s,q)=\psi(s)\Phi(q)$,
this equation becomes
$$
\frac{i{\partial \psi(s)\over\partial s}}{\psi(s)}=\lambda=
\frac{\hat{\cal H}\Phi(q)}{\Phi(q)}.
\EQN cv.3$$
For $\psi(s)\propto exp(-ims/2)$, one finds
$$
\hat{\cal H}\Phi(q)={(\hat p+eA)^2\over 2m}\Phi(q)={m\over 2}\Phi(q), 
\EQN cv.4$$
with $\hat p^\mu=i\partial ^\mu$.
For the free particle case, $A^\mu=0$, 
$$
\hat{\cal H}={(\hat p)^2\over 2m}=-{1\over 2m} \square^2,
\EQN cv.6$$
with 
$$
\square^2={\partial ^2 \over \partial ^2 q^0}-\nabla^2=
{\partial ^2 \over \partial ^2 q^0}-{\partial ^2 \over \partial ^2q^1}
-{\partial ^2 \over \partial ^2 q^2}-{\partial ^2 \over \partial ^2 q^3}.
\EQN cv.7$$
Potential problems can be included using the vector potential
$A^\mu=(-V(q),0,0,0)$, and the equation for $\Phi(q)$
becomes
$$
((\hat p^0-V(q))^2+\nabla^2)\Phi(q)=m^2\Phi(q).
\EQN cv.8$$
The time dependence is this equation is separable, and the solution
$\Phi(q)$ takes the form
$$
\Phi(q)\propto e^{-iEt}\phi(x,y,z).
\EQN cv.9$$
This gives 
$$
((E-V(x,y,z))^2+\nabla^2)\phi(x,y,z)=m^2\phi(x,y,z).
\EQN cv.10$$
In the one dimensional case, this becomes
$$
(E-V(z))^2\phi(z)=m^2\phi(z)-{\partial^2\phi(z) \over \partial^2 z}.
\EQN cv.11$$
The non-relativistic limit is found from
$$
(E-V(z))\sqrt{\phi(z)}=m\sqrt{\phi(z)}
\sqrt{1-\frac{1}{m^2\phi(z)}\frac{d^2\phi(z)}{dz^2}},
\EQN cv.12$$
and this becomes
$$
(E-V(z))\sqrt{\phi(z)}
\approx m\sqrt{\phi(z)}(1-\frac{1}{2m^2\phi(z)}\frac{d^2\phi(z)}{dz^2}+\dots),
\EQN cv.13$$
which is equivalent to
$$
\frac{d^2\phi(z)}{dz^2}+2m(E-m-V(z))\phi(z).
\EQN cv.14$$
This is the usual time-independent Schr\"odinger equation;
however, the eigenvalues $E-m$ contain the rest mass of the particle. The
rest mass should be retained in the non-relativistic limit because in the free
particle case the phase of the wave function is a Lorentz invariant, 
$Et-\vec p\cdot\vec q=ms$. Furthermore, the operator $\hat p^0$ is a
generator for displacement of the time coordinate $q^0$; however, in this
theory, it is not associated with a phase space Hamiltonian operator through a
time dependent Schr\"odinger equation. In this way, time and energy are not
conjugate coordinates expressible in terms of phase space operators, and an
uncertainty relation of the type described in Appendix A is not possible. A
related comment can be found in \cite{[WP]}.
\par
The equation of continuity for a free scalar particle follows from the 
covariant Schr\"odinger type equation.
From \Eq{cv.2} and the fact that ${\hat{\cal H}}$ is Hermitian, one can form
$$
\psi^*(q,s)i\frac{ \partial\psi(q,s)}{\partial s}
+\psi(q,s)i\frac{\partial\psi^*(q,s)}{\partial s}
=\psi^*(q,s){\cal H}\psi(q,s)-\psi(q,s){\cal H}\psi^*(q,s),
\EQN cv.15$$
and this can be written as
$$
i\frac{\partial \rho(q,s)}{\partial s}-\hat p_\mu j^\mu(q,s)=0,
\EQN cv.16$$
with 
$$
\rho(q,s)=\psi^*(q,s)\psi(q,s),
\EQN cv.17$$
and
$$
j^\mu(q,s)=\frac{1}{2m}
(\psi^*(q,s)\hat p^\mu\psi(q,s)-\psi(q,s)\hat p^\mu\psi^*(q,s)).
\EQN cv.18$$
Here $\rho(q,s)$ represents the probability density in the 
four-vector $q^\mu$ representation, and $j^\mu(q,s)$ is associated with 
the charge density current\cite{[PandW]}.
The normalization condition $\scp{\Psi(s)}{\Psi(s)}=1$ requires
$$
\int d^4q\,\rho(q,s)=1.
\EQN cv.19$$
For this theory, the probability density is positive definite, and 
this removes the historical problem associated with finding a suitable
probability density for the Klein-Gordon equation.

\section{FREE PARTICLE GREEN'S FUNCTION}

\par
The evolution of a free particle quantum state is determined from the
the free particle Green's function.
The Green's function for the covariant scalar quantum state $\ket{\Psi(s)}$ is 
defined as
$$
g(q,q',s)=\theta(s)G(q,q',s)=\theta(s)\bra{q}U^{-s{\hat{\cal H}}}\ket{q'},
\EQN cvgf.1$$
where $\theta(s)$ is the Heaviside function, and it satisfies the equation
$$
(i\frac{\partial}{\partial s}-{\hat{\cal H}})g(q,q',s)
=\delta(s)\delta^4(q-q'),
\EQN cvgf.2$$
and the condition
$$
\lim_{s\to 0}g(q,q',s)=\delta^4(q-q').
\EQN cvgf.3$$
Following the method of Dirac \cite{[Dirac33]}, it can be obtained from 
the classical covariant action as
$$
g(q,q',s)=\theta(s)A(s)e^{i{{\cal S}(q,q',s)}}.
\EQN cvgf.4$$
For the free particle covariant Hamiltonian ${\hat{\cal H}}
=\hat p\cdot\hat p/(2m)$, the solutions to the classical equations of motion
are
$$
q^\mu(s)-{q^\prime}^\mu(s)=sp^\mu(s)/m,\,\, {\rm and} \,\, p^\mu(s)
={\rm constant},
\EQN cvgf.5$$ 
and the covariant classical action
becomes
$$
{\cal S}(q,q',s)=\int\frac{m}{2}\frac{dq}{ds}\cdot\frac{dq}{ds}ds
=\frac{m(q-q')\cdot(q-q')}{2s}. 
\EQN cvgf.6$$
The Green's function that satisfies \Eq{cvgf.2} and \Eq{cvgf.3} is
$$
g(q,q',s)=\theta(s)(\sqrt{\frac{m}{i2\pi s}})^4
e^{i\frac{m}{2s}[(t-t')^2-(x-x')^2-(y-y')^2-(z-z')^2]}.
\EQN cvgf.7$$
\par This evolution function can be used to find the wave function $\psi(q,s)$
from the initial state $\psi(q,0)$ using
$$
\psi(q,s)=\int G(q,q',s)\psi(q',0)d^4q'.
\EQN cvgf.8$$
Three cases of free particle motion are considered:

\noindent {\it case 1}: The initial particle is localized in space-time, with 
wave function 
$$
\psi(q,0)=\delta(q^0)\delta(q^1)\delta(q^2)\delta(q^3),
\EQN cvgf.9$$
and the evolved wave function is
$$
\psi(q,s)=(\sqrt{\frac{m}{i2\pi s}})^4
e^{i\frac{m}{2s}[(t)^2-(x)^2-(y)^2-(z)^2]}.
\EQN cvgf.10$$
\noindent{\it case 2}: The initial particle is delocalized in space-time, with 
wave function 
$$
\psi(q,0)=\frac{1}{4\pi^2}e^{-ip\cdot q}=\frac{1}{4\pi^2}e^{-ip^0q^0+i\vec p\cdot\vec q},
\EQN cvgf.11$$
and the evolved wave function is
$$
\psi(q,s)=\frac{1}{4\pi^2}e^{-ip\cdot q}e^{i(-p^0p^0+\vec p\cdot\vec p)s/(2m)}.
\EQN cvgf.12$$
\noindent{\it case 3}: The initial particle is described by 
Gaussian distributions, with initial state wave function
$$\eqalign{
\psi(q,0)&=\frac{1}{2\pi (\sigma_0\sigma_1\sigma_2\sigma_3)^{1/2}}
e^{-ip(0)\cdot q}\cr
\times&e^{-(q^0)^2/(4\sigma_0^2)}
e^{-(q^1)^2/(4\sigma_1^2)}\cr
\times&e^{-(q^2)^2/(4\sigma_2^2)}
e^{-(q^3)^2/(4\sigma_3^2)},\cr
}
\EQN cvgf.13$$
and the evolved wave function is
$$\eqalign{
\psi(q,s)&=e^{i\tan^{-1}(s)}
\psi_0(q^0,p^0(0),s)
\psi_1(q^1,-p^1(0),-s)\cr
&\times\psi_2(q^2,-p^2(0),-s)
\psi_3(q^3,-p^3(0),-s),\cr
}
\EQN cvgf.14$$
with 
$$\eqalign{
\psi_\mu(q^\mu,p^{\mu}(0),s)&=\frac{1}{(2\pi\sigma^2_\mu(s))^{1/4}}
e^{-\frac{(q^\mu-p^\mu(0)s)^2}{2\sigma^2(s)}(1+is)}\cr
&\times e^{i\frac{(p^\mu(0))^2s}{2}}
e^{-iq^\mu p^\mu(0)},\cr
}
\EQN cvgf.15$$
where
$$
\sigma_\mu(s)=\sigma_\mu\sigma(s)=\sigma_\mu\sqrt{1+s^2},
$$
and where
$$\eqalign{
q^\mu&\rightarrow q^\mu/\delta(q^\mu)\cr
p^\mu&\rightarrow p^\mu/\delta(p^\mu)\cr
s&\rightarrow s/\delta(s)\cr
\delta(q^\mu)\delta(p^\mu)&=\hbar\cr
\delta(s)\delta(p^\mu)&=mc\delta(q^\mu)\cr
\delta(q^\mu)&=\sqrt{2}\sigma_\mu.\cr
}
\EQN cvgf.16$$
This is the free particle motion, which is to be compared with the classical
motion resulting from  \Eq{pp.6} and \Eq{pp.7}.

\par
Each of these solutions can be confirmed to be a solution of the
covariant Schr\"odinger type equation \Eq{cv.2}. In addition each is a solution to
the equation of continuity \Eq{cv.16}. The normalization condition \Eq{cv.19}
is satisfied by the wave function \Eq{cvgf.14}. The wave function
\Eq{cvgf.9} must be viewed as the limit as $\sigma_\mu \rightarrow 0$
in \Eq{cvgf.13}. However; \Eq{cvgf.11} satisfies the condition for 
delta function normalization.
The function $\psi^*(q,-s)$ are also solutions to the covariant
Schr\"odinger type equation and the equation of continuity. They represent 
the solutions obtained from the initial functions above with $p\rightarrow
-p$. This is seen from the complex conjugation of the covariant
Schr\"odinger type equation followed by $s\rightarrow -s$. Also for
the free particle Green's function 
$$
g(q,q',s)=g^*(q,q'-s),
\EQN cvgf.17$$
and
$$
\psi^*(q,-s)=\int g^*(q,q',-s)\psi^*(q',0)d^4q.
\EQN cvgf.18$$

\par
The appropriate wave functions to associate with the classical limits of 
a quantum theory are found from the minimum uncertainty states. For the spatial
coordinates, these states are found from the eigenvalue equation
$$
\hat a^j\ket{a^j}=\frac{\hat q^j+i\hat p^j}{\sqrt{2}}\ket{a^j}=a^j\ket{a^j},
\EQN mus.1$$
for $j=1,2$ or $3$ with $a^j=(q^j(0)+ip^j(0))/\sqrt{2}$ and 
$[\hat q^i,\hat p^j]=i\delta_{ij}$. The normalized 
wave functions that satisfy this equation
are

$$
\scp{q^i}{a^i}=\frac{1}{\pi^{1/4}}e^{ip^i(0)q^i-(q^i-q^i(0))^2/2}.
\EQN mus.2$$
For the time component, the coordinate and momentum space  commutation 
relations are
$$
[q^0,i\frac{\partial}{\partial q^0}]=-i,\quad
[p^0,-i\frac{\partial}{\partial p^0}]=i.
\EQN mus.3$$
The momentum space minimum uncertainty state is found from
$$
\hat b^0\ket{b^0}=\frac{\hat p^0+i\hat q^0}{\sqrt{2}}\ket{b^0}=b^0\ket{a^0},
\EQN mus.4$$
with $b^0=(p^0(0)+iq^0(0))/\sqrt{2}$, and the wave function
is
$$
\scp{p^0}{b^0}=\frac{1}{\pi^{1/4}}e^{iq^0(0)p^0-(p^0-p^0(0))^2/2}.
\EQN mus.5$$
The coordinate space function is found from
$$
\scp{q^0}{b^0}=\int dp^0\scp{q^0}{p^0}\scp{p^0}{b^0},
\EQN mus.6$$
with $\scp{q^0}{p^0}=(1/\sqrt{2\pi})\exp(-ip^0q^0)$, and it becomes
$$
\scp{q^0}{b^0}=\frac{1}{\pi^{1/4}}e^{-ip^0(0)(q^0-q^0(0))-(q^0-q^0(0))^2/2}.
\EQN mus.7$$
The momentum space functions associated with the coordinate space functions
are found from the complex conjugate of \Eq{mus.2} and the replacement
$q \leftrightarrow p$.
The uncertainty for an  operator $\hat Q$ is  defined with respect to the
state $\ket{\psi}$ as 
$$
\sigma(Q)
=\sqrt{\bra{\psi}(\hat Q-\bar Q)^2\ket{\psi}},
\EQN mus.8$$
with $\bar Q=\bra{\psi}\hat Q\ket{\psi},$
and, for the minimum uncertainty states, they have the value
$\sigma(q^\mu)=\sigma(p^\mu)=\frac{1}{\sqrt{2}}$.
It is these states that are used for the initial state wave functions
that appear in \Eq{cvgf.13}, which are found with the coordinate replacements
\Eq{cvgf.16}.
\par
The classical representations of the parameters $m$ and $s$
are
$$
m=\sqrt{p\cdot p},\quad {\rm and}\quad s=\sqrt{q\cdot q}.
\EQN dqm.3$$
The Poisson bracket of these quantities has the value
$$
\{m,s\}=\frac{p.q}{sm}=1,
\EQN dqm.4$$
and this suggest that there could be a  corresponding commutation relation 
for the 
operators associated with $m(p)$ and $s(q)$. However, the spectrum of $m$
must be positive, and as shown in Appendix A this is not possible if 
$[\hat m,\hat s]=i$.

\section{PARTICLE IN AN EXTERNAL FIELD}

\par
To illustrate the method of solution for \Eq{cv.8}, I consider the relativistic 
motion of a particle in an 
external field with potential $A^0(z)=-V(z)=Kz$. This is the covariant
quantum analogue to the classical example given in \Eq{pp.8}.
The equation for $\Phi(z)$ becomes
$$
(E+kz)^2\phi(z)+\frac{\partial^2}{\partial z^2}\phi(z)=m^2\phi(z).
\EQN exf.1$$
Using the substitutions
$$
\xi=\frac{1}{\sqrt{k}}(E+kz),\quad \lambda=-\frac{m^2}{k},
\EQN exf.2$$
one finds
the parabolic cylindrical equation \cite{[GandR]}
$$
{d^{2}\phi(\xi)\over d\xi^{2}}+(\xi^{2}+\lambda)\phi(\xi)=0,
\EQN exf.3$$
$$
\phi(\xi)=D_{-{1+i\lambda\over2}}[\pm(1+i)\xi],
\EQN exf.4$$
$$
{\quad D_{p}[(1+i)\xi]={2^{{p+1\over2}}\over \Gamma\left(-{p\over2}
\right)}{\int_{1}^{\infty}}e^{-{i\over2}\xi^{2}x}{(x+1)^{{p -1\over2}}\over
(x-1)^{1+{p\over2}}}  dx},
\EQN exf.5$$
where for $[{\rm RE}\, p< 0;\,{\rm RE}\, i\xi^{2}\geq0]$,
$$
\def\UUU{\quad {D_{p}(\xi)}}

\eqalign{\UUU&={2^{{1\over4}+{p\over2}}W_{{1\over4}+{p\over2},-{1\over4}}
\left({\xi^{2}\over2}\right)\xi^{-{1\over2}}{}}\hfill\cr
&={2^{{p\over2}}e^{-{\xi^{2}\over4}}\left\lbrace {\sqrt{\pi}\over  \Gamma
\left({1-p\over2}\right)}\Phi \left(-{p\over2},   {1\over2};   {\xi^{2}
\over2}\right)-{\sqrt{2\pi}\xi\over \Gamma\left(-{p\over2}\right)}\Phi 
\left({1-p\over2},   {3\over2};   {\xi^{2}\over2}\right)\right\rbrace },\hfill\cr}
\EQN exf.6$$
and where $W_{\lambda,\mu}(z)$ are Whittaker functions, and
$\Phi(\alpha,\gamma;x)$ are confluent hypergeometric functions.
The eigenvalue spectrum for the energy is continuous, and in the
non-relativistic limit the wave functions become Airy functions.

\section{COVARIANT FERMION CASE}

\par A similar analysis can be done for the case of spin $1/2$ particles.
The covariant Hamiltonian associated with the classical equations is
$$
{\cal H}=\frac{\bra{\bar u}\ps+e\As(q)\ket{u}}{2m}
=(p+eA)_\mu\frac{\bra{\bar u}\gamma^\mu\ket{u}}{2m},
\EQN fc.1$$
where $\ket{u}$ is a spinor and $\bra{\bar u}=\bra{u}\gamma^0$ with
normalization $\scp{\bar u}{u}=2\sqrt{p\cdot p}$. For scalar products and
matrix elements with spinors, an average is made over helicity states.
Hamilton's equations for this case give
$$
\frac{dq^\mu}{ds}=\frac{\partial{\cal H}}{\partial p_\mu}=\frac{\bra{\bar u}\gamma^\mu\ket{u}}{2m}
\EQN fc.2$$
$$
\frac{dp^\mu}{ds}=-\frac{\partial{\cal H}}{\partial q_\mu}
=-\frac{\partial eA_\nu}{\partial q_\mu}\frac{\bra{\bar
u}\gamma^\nu\ket{u}}{2m}=-\frac{\partial eA\cdot u}{\partial q_\mu},
\EQN fc.3$$
and this gives \Eq{ac.5}.
\par
The covariant wave equation associated with this case is
$$
i\frac{\partial \ket{\psi(s)}}{\partial s}=(i\hat\ps+e\As)\ket{\psi(s)}
=\hat{\cal H}\ket{\psi(s)},
\EQN fc.4$$
and the Hamiltonian is
$$
\hat{\cal H}=(\hat\ps+e\As)=\gamma^0{\hat{\cal H}}^\dagger\gamma^0.
\EQN fc.5$$

The state $\ket{\psi(s)}$ is generated from
$\ket{\psi(0)}$ using
$$
\ket{\psi(s)}=\hat U(s)\ket{\psi(0)}=e^{-is{\hat{\cal H}}}\ket{\psi(0)}
\EQN fc.7$$
$$
\bra{\bar\psi(s)}=\bra{\bar\psi(0)}\hat U(-s)=\bra{\bar\psi(0)}e^{is{\hat{\cal H}}},
\EQN fc.8$$
such that
$$
\hat U(-s)=\gamma^0\hat U^\dagger(s)\gamma^0\quad {\rm and}\quad
\scp{\bar\psi(s)}{\psi(s)}=1.
\EQN fc.9$$
\par
In the $q$-representation, the probability density for the four-vector $q$
is given by
$$
\rho(q,s)=\scp{\bar{\psi}(s)}{q}\scp{q}{\psi(s)},
\EQN fc.11$$
which is normalized such that
$$\eqalign{
\int \rho(q,s)d^4q&=\scp{\bar{\psi}(s)}{\psi(s)}=\bra{\bar{\psi}(0)}
{\hat U}^{-1}(s){\hat U}(s)\ket{\psi(0)}\cr
&=\scp{{\bar \psi}(0)}{\psi(0)}=1.\cr
}
\EQN fc.12$$
Pre-multiplying the covariant equation \Eq{fc.4} for $\scp{q}{\psi(s)}$ by
$\scp{{\bar\psi}(s)}{q}$, and post-multiplying the matrix transformed complex 
conjugate of \Eq{fc.4}
by $\gamma^0\scp{q}{\psi(s)}$ gives, when the second resulting equation is
subtracted from the first, 
the equation of continuity
$$
\frac{\partial \rho(q,s)}{\partial s}=
\frac{\partial j^\mu(q,s)}{\partial q^\mu},
\EQN fc.13$$
where the current density is
$$
j^\mu(q,s)=\scp{{\bar\psi}(s)}{q}\gamma^\mu\scp{q}{\psi(s)}.
\EQN fc.14$$
Upon integration over the four-volume $d^4q$, one finds from \Eq{fc.13}
$$
\frac{\partial\int \rho(q,s)d^4q}{\partial s}=
\int\frac{\partial j^\mu(q,s)}{\partial q^\mu}d^4q=\int j^\mu(q,s)dS_\mu=0,
\EQN fc.15$$
where $dS_\mu$ is an element of three dimensional hypersurface orthogonal
to the direction $\mu$. This is the statement of conservation of charge.
As an example, I consider the free particle case when the spinor state is 
expanded in
four-momentum $p$ and helicity $\lambda$ states such that
$$\eqalign{
\ket{\psi(s)}
&=\sum_\lambda\int \hat
U(s)a_\lambda(p)\frac{\ket{u(p,\lambda)}\otimes\ket{p}}{\sqrt{\scp{\bar
u(p,\lambda)}{u(p,\lambda)}}}d^4p\cr
&=\sum_\lambda\int
e^{-is\sqrt{p.p}}a_\lambda(p)\frac{\ket{u(p,\lambda)}\otimes\ket{p}}{  
\sqrt{\scp{\bar u(p,\lambda)}{u(p,\lambda)}}}d^4p,\cr
}
\EQN fc.16$$
and the wave function in the $q$ representation is 
$$
\scp{q}{\psi(s)}
=\sum_\lambda\int e^{-is\sqrt{p.p}}e^{-ip\cdot q}
\frac{a_\lambda(p)\ket{u(p,\lambda)}}
{\sqrt{\scp{\bar u(p,\lambda)}{u(p,\lambda)}}}
\frac{d^4p}{(\sqrt{2\pi})^4}.
\EQN fc.17$$
This leads to the density function
$$\eqalign{
\rho(q,s)&=\cr
&\sum_\lambda\sum_{\lambda'}\int d^4p'\int
d^4p\,a^*_\lambda(p')a_\lambda(p) e^{-is(\sqrt{p.p}-\sqrt{p'.p'})}e^{-i(p-p')\cdot q}\cr
&\times \frac{\scp{\bar
u(p',\lambda')}{u(p,\lambda)}}
{\sqrt{\scp{\bar u(p',\lambda')}{u(p',\lambda')}}
\sqrt{\scp{\bar u(p,\lambda)}{u(p,\lambda)}}}
\frac{1}{(\sqrt{2\pi})^8},\cr
}
\EQN fc.18$$
and the charge current density becomes
$$
j^\mu(q,s)=e\sum_\lambda\int\vert a_\lambda(p)\vert^2 
\frac{\bra{\bar u(p,\lambda)}\gamma^\mu\ket{u(p,\lambda)}}
{\scp{\bar u(p,\lambda)}{u(p,\lambda)}}d^4p,
\EQN fc.19$$
or
$$
j^\mu(q,s)\approx e\sum_\lambda\int\vert a_\lambda(p)\vert^2\, 
\frac{p^\mu}{m}d^4p.
\EQN fc.20$$

\section{PARAMETER VALUES AND NORMALIZATION}

\par
The parameters $m$ and $s$ that appear in the covariant  evolution equations
\Eq{cqmsb.5}, and in the classical equations are associated with
quantum operators. This association
is described for the mass parameter, and the corresponding relation for the
proper time parameter $s$ may be found is a similar manner.
For the spin zero scalar case, the quantum state for a free particle may be expanded
in terms of the four-momentum eigenstates as
$$
\ket{\Psi(s)}=\int  a_p(s)\,\ket{p}\,d^4p,
\EQN n.1$$
and the scalar product is
$$
1=\scp{\Psi(s)}{\Psi(s)}=\int \vert a_p(0)\,\vert^2\,d^4p.
\EQN n.2$$
The mass parameter $m$ is found from
$$
m^2=\bra{\Psi(s)}\hat p\cdot \hat p\ket{\Psi(s)}
=\int \vert a_p(0)\vert^2 p\cdot pd^4p.
\EQN n.4$$
\par
For the spin $1/2$ case, the spinor state is
$$
\ket{\Psi(s)}=\sum_\lambda\int
a_\lambda(p,s)\,\frac{\ket{u(p,\lambda)}\otimes\ket{p}}
{\sqrt{\scp{\bar u(p,\lambda)}{u(p,\lambda)}}}\,d^4p,
\EQN n.6$$
where $\ket{u(p,\lambda)}$ is a spinor that is normalized such that
$$
\scp{\bar u(\pm p,\lambda)}{u(\pm p,\lambda)}=\pm 2\sqrt{p\cdot p},
\EQN n.7$$
and satisfies the condition
$$
\bra{\bar u(\pm p,\lambda)}\gamma^\mu\ket{u(\pm p,\lambda)}=2p^\mu.
\EQN n.8$$
The scalar product in this case is
$$
\scp{\bar \Psi(s)}{\Psi(s)}
=\sum_\lambda\int \vert a_\lambda(p)\,\vert^2\,d^4p=1.
\EQN n.9$$
The mean value of the operator $\hat \ps$ is 
$$
\frac{\bra{\bar\Psi(s)}{\hat \ps}\ket{\Psi(s)}}{\scp{\bar \Psi}{\Psi}}
=\sum_\lambda {\int \vert a_\lambda(p)\vert^2 \sqrt{p\cdot p}\, d^4p}.
\EQN n.10$$
\par
In general, the moments of the operators $\hat {\cal H}$ 
may be evaluated, and these can be used to construct the associated
characteristic function. The distributions associated with this operator
can be found from the Fourier transforms of the characteristic function.
For example, the characteristic function for the operator $\hat {\cal H}$ is
$$
\Phi(\alpha)=\bra{\Psi(s)}e^{i\alpha\hat {\cal H}}\ket{\Psi(s)},
\EQN n.11$$
and the distribution of $m/2$ is
$$
\rho(m/2,s)=\frac{1}{\sqrt{2\pi}}\int_{-\infty}^\infty
e^{-i\alpha m/2}\Phi(\alpha)d\alpha.
\EQN n.12$$
This shows the quantum nature of the mass distribution, which in
principle can be measured. For example, measurements of the particle speed of
an ensemble of identical free particles with identical initial conditions
can be made. Since the magnitude of the three momentum and energy are related
by the relation $\vert \vec p\vert=\beta E$, the value of $m^2$
can be determined with an independent measurement of either $\vert \vec
p\vert$ or $E$. For a neutral particle, $E$ can be found from an impact;
however, for a charged particle, $\vert \vec p\vert$ can be found from its
curvature in a magnetic field. For an ensemble of free particles, 
measurements of $s^2$ can be made from the
observation of the distance traveled by a free particle in time $t$  and the
measurement of the distance $q^0=ct$ traveled by light in the same time.

\section{FOUR-VELOCITY AND FOUR-ACCELERATION OPERATORS}

\par  The observed values of four-velocity are 
associated with the four-velocity operator $\hat v^\mu=i[\hat q^\mu,\hat{\cal H}]$.
These values are found in 
the scalar case, with $\hat{\cal H}$ given by \Eq{cv.4}, from the expectation value 
$$
\frac{dq^\mu}{ds}= u^\mu =\frac{\bra{\Psi(s)}i[\hat q^\mu,\hat{\cal H}]\ket{\Psi(s)}}
{\scp{\Psi}{\Psi}},
\EQN fv.1$$
and in the spinor case from
$$
\frac{dq^\mu}{ds}=u^\mu =\frac{\bra{\bar \Psi(s)}i[\hat q^\mu,{\hat {\cal
H}}]\ket{\Psi(s)}}
{\scp{\bar\Psi}{\Psi}},
\EQN fv.2$$
where $\hat{\cal H}$ is given by \Eq{fc.5}.
In the free particle case, these 
both become
$$
 u^\mu =\frac{p^\mu}{m},
\EQN fv.3$$
when the phase space operators satisfy the quantum bracket condition
$$
[{\hat q}^\mu,{\hat p}^\nu]=-ig^{\mu\nu}.
\EQN fv.4$$
The usual velocity vectors are found from $dq^i/dq^0$. It is clear
that the operators associated with four-velocity are unambiguously defined,
and that the matrix elements in \Eq{fv.1} and \Eq{fv.2} give the predicted observed
mean values for four-velocity. The operator $\hat {\cal H}$ is either linear or 
quadratic in $\hat p^\mu$, and it does not involve the square root of the
scalar product of the three momentum operator. This is a clear advantage over the 
non-covariant forms in 
\cite{[A-WIG]} and \cite{[OC-WIG]} proposed to represent the velocity operator.
\par 
From the four-velocity operator, one can also obtain the quantum equations
that correspond to the classical equations of motion for a point
particle of mass $m$ interacting with a four-vector potential. For the
scalar case, the four-velocity operator $\hat u^\mu=(\hat p^\mu+e A^\mu)/m$
satisfies the commutation relation
$$
[\hat u^\mu,\hat u^\nu]=\frac{ie}{m^2}F^{\mu\nu},
\EQN fv.5$$
and the four-acceleration operator is found from
$$
\hat a^\mu=i[\hat u^\mu,\hat{\cal H}],
\EQN fv.6$$
which gives
$$
ma^\mu+\frac{e}{2}\bra{\Psi(s)}(F^{\mu\nu}\hat u_\nu+\hat
u_\nu F^{\mu\nu})\ket{\Psi(s)}, 
\EQN fv.7$$
which is to be compared with \Eq{ac.7}. For the spinor case with interaction, 
$$
u^\mu=\frac{\bra{\bar\Psi(s)}\gamma^\mu\ket{\Psi(s)}}{\scp{\bar
\Psi}{\Psi}},
\EQN fv.8$$
and 
$$
\frac{dp^\mu}{ds}=\frac{\bra{\bar\Psi(s)} i[\hat p^\mu,\hat \ps+e\As]\ket{\Psi(s)}}{\scp{\bar
\Psi}{\Psi}}.
\EQN fv.9$$
This becomes
$$
\frac{dp^\mu}{ds}+e\frac{\bra{\bar\Psi(s)}\partial^\mu\As\ket{\Psi(s)}}{\scp{\bar
\Psi}{\Psi}},
\EQN fv.10$$
which is to be compared with \Eq{ac.5}.
It is important to note that consistency of these equations with the classical
results depends upon the
commutation relation \Eq{fv.4}, and this justifies its introduction.

\section{CONCLUSIONS}

\par
We have seen that the covariant formulation of quantum mechanics provides
representations that include both the scalar and spinor description of
particle states. It gives a clear connection of the covariant quantum
description with classical covariant particle dynamics. The covariant free
particle Green's function describes the evolution of initial states in terms
of the proper time parameter $s$, and it is seen how the initial
Gaussian state propagates with spreading width with respect to the classical
covariant path. In addition, it gives an acceptable representation of the
probability density in both the scalar and spinor representations, and it
produces a new set of equations of continuity, which show the connection
between probability density, charge density, and current density.
The solution of potential problems also
follows from this formulation. This permits the description of practical
relativistic quantum effects, which are the relativistic generalization of
standard non-relativistic quantum situations. Also in the covariant
formulation, the representation of the four-velocity and the
four-acceleration operators is
unambiguous, and one can clearly see how these operators are associated with
observed values. 
\par The properties of quantum amplitudes are shown to be associated with
an Hamiltonian formulation, which acts on the space of these amplitudes.
This leads to the notion of invariance of the symmetric bracket, which is
defined on this space. This invariance and the associated Hamiltonian
formulation are a consequence of the covariant Schr\"odinger type equation. This
provides a subtle connection between the probability interpretation of
quantum theory and the evolution of quantum states. It is shown in
\cite{[TG-SK]} how the the non-covariant form of this symmetric bracket
invariance leads to second quantization for fermion and boson states.
\vfill\eject
\leftline{REFERENCES}
\medskip
\hrule
\medskip
\ListReferences
\vfill\eject
\appendix{A}{MINIMUM UNCERTAINTY PRINCIPLE}

\par
For the Hermitian operators $\hat A$ and $\hat B$ that satisfy the commutation
relation $[\hat A, \hat B]=i$, one defines
the deviation operators $\hat \Delta(A)$ and $\hat \Delta(B)$ where
$\hat \Delta(Q)=\hat Q-\bar Q$,  
with $\bar  Q=\bra{\psi}\hat Q\ket{\psi}$. It follows that
$[\hat \Delta(A), \hat \Delta(B)]=i$. 
Introducing the operator
$$
\hat a=\frac{\hat A+i\lambda\hat B}{\sqrt{2}},
\EQN mup.1$$
for real $\lambda$,
the inner product inequality
$$
\bra{\psi}{\hat \Delta}^\dagger(a)\hat \Delta(a)\ket{\psi}\geq 0
\EQN mup.2$$
becomes
$$
(\lambda -\frac{1}{2\sigma^2(B)})^2+\frac{\sigma^2(A)}{\sigma^2(B)}\geq
\frac{1}{4\sigma^4(B)},
\EQN mup.3$$
where $\sigma(Q)$ is defined in \Eq{mus.8}.
This equation implies 
$$
\sigma(A)\sigma(B)\geq \frac{1}{2}.
\EQN mup.4$$
If equality is satisfied in \Eq{mup.2} and \Eq{mup.4}, then
$\lambda=1/(2\sigma^2(B)$. The quantum state closest to the classical
result has $\sigma(A)=\sigma(B)$, and this implies $\lambda =1$.
These conditions imply
$$
\hat \Delta(a)\ket{a}=0,
\EQN mup.5$$
where $\ket{a}$ is the well known coherent state, which is generated from the
vacuum state $\ket{0}$ with the displacement operator
$$
\ket{a}=\hat D(a)\ket{0}=e^{(a\ad-a^*\ah)}\ket{0}.
\EQN mup.6$$
\par
If $\hat A$ and $\hat B$ are Hermitian and satisfy the 
commutation relation $[\hat A, \hat B]=i$, then the spectrum of $\hat B$ is in general
both positive and negative and unbounded in the $B$ 
representation where $\hat B\ket{B}=B\ket{B}$. This is seen
from
$$
\hat U(A)\hat B\hat U^\dagger(A)\hat U(A)\ket{B}=B\hat U(A)\ket{B},
\EQN mup.7$$
which become
$$
\hat B\hat U(A)\ket{B}=(B+\phi)\hat U(A)\ket{B},
\EQN mup.8$$
where $\hat U(A)=\exp(i\phi\hat A)$, with $-\infty <\phi<\infty$. A similar
result is found for the $A$ representation.

\bye